\documentclass[sigconf,nonacm]{acmart}

\IfFileExists{libertine.sty}{}{\usepackage{lmodern}}
\usepackage{booktabs}
\usepackage{array}
\usepackage[ruled,vlined,linesnumbered]{algorithm2e}
\usepackage{placeins}
\usepackage{tikz}
\usepackage{xcolor}
\usepackage{xspace}
\usetikzlibrary{arrows.meta}
\setcopyright{none}
\renewcommand\footnotetextcopyrightpermission[1]{}
\acmConference[Preprint]{Technical report}{August 2026}{}

\newcommand{\system}{SCOUT\xspace}
\newcommand{\gemini}{GEMINI\xspace}

\definecolor{revisionblue}{RGB}{0,82,155}

\DontPrintSemicolon
\SetKwProg{Fn}{Function}{:}{end}
\SetKw{Return}{return}

\title{\system: Symmetric Consensus Outlier Detection for Failure Localization in LLM Pre-Training}
\author{Zhuang Wang}
\authornote{Independent Researcher; \texttt{wangzhuangowl@gmail.com}.}

\begin{document}

\begin{abstract}
In LLM pre-training, synchronization propagates rank-local stalls, slowdowns, and numerical errors into job-wide symptoms, obscuring their origin.
Existing diagnosis often relies on in-process monitors that cannot report after the trainer blocks or terminates, or on post-mortem logs that preserve only synchronized symptoms; offline health tests lose the workload and operating conditions that triggered the failure.
We present \system, a unified runtime failure-localization framework built on one design principle: identify outliers through strict-majority consensus among equivalent replicas.
\system aligns replica progress, timing, and numerical evidence, then uses its \emph{Consensus Collective Communication} (C3) abstraction to identify ranks whose compact signatures disagree with their peers.
An out-of-band CPU observer remains responsive when training hangs, whereas in-situ replay exercises recurring stragglers and silent data corruption (SDC) beside the live job with its model state, kernels, allocations, communication path, and thermal and memory pressure present.
Collective fingerprints expose rank-local protocol divergence.
Clean replay coverage certifies checkpoint numerical integrity, preventing recovery from selecting state corrupted by SDC.
\system integrates with PyTorch, TorchTitan, Megatron-Core, and DeepSpeed without training-loop or framework-source modifications.
\system is open source at \url{https://github.com/LMResiliency/lm-resiliency}.
\end{abstract}

\maketitle
\pagestyle{plain}

\section{Introduction}

Modern LLM pre-training is a distributed systems workload spanning tens of thousands of accelerators.
MegaScale reports training a 175-billion-parameter model on 12,288 GPUs~\cite{megascale-nsdi24}.
Meta reports pre-training the nearly two-trillion-parameter Llama~4 Behemoth with 32,000 GPUs~\cite{llama4-meta25}.
Because model and optimizer state exceed one accelerator's memory, training distributes them across worker processes, or \emph{ranks}.
Model and optimizer state are partitioned across ranks with data parallelism~\cite{pytorch-ddp-pvldb20}, tensor parallelism~\cite{megatron-arxiv19}, pipeline parallelism~\cite{gpipe-neurips19}, expert parallelism~\cite{gshard-arxiv20}, and state sharding~\cite{fsdp-pvldb23,zero-sc20}.
These hybrid-parallel jobs are typically synchronous: within every optimization iteration, ranks exchange tensors through collective communication and wait at data or control dependencies before advancing~\cite{megatron-arxiv19,zero-sc20,megascale-nsdi24}.

At this scale, reliability directly determines delivered throughput because one delayed or failed component can idle many healthy accelerators.
While a synchronous job waits for diagnosis or rollback, its reserved accelerators continue accruing GPU-hours without completing training steps.
An $\mathcal{O}(10{,}000)$-GPU cluster recorded 100--230 critical failures per week~\cite{aegis-nsdi25}, and another large distributed-training environment observed up to two faults per day for an individual job using thousands of machines~\cite{minder-nsdi25}.
Performance faults are comparably costly: a five-month trace found stragglers in 42.5\% of jobs, wasting 10.4\% of allocated GPU-hours~\cite{stragglers-osdi25}, while a separate production study found fail-slows delaying large jobs by 1.34$\times$ on average~\cite{greyhound-atc25}.

Detecting a failure is insufficient to resume training because recovery must also localize where to act.
A timeout, slowdown, or numerical anomaly does not say which host or communication path to replace or which checkpoint to restore.
A wrong decision can restart the job on the same faulty resource, quarantine a healthy one, or restore corrupted state~\cite{minder-nsdi25,aegis-nsdi25}.
Minder~\cite{minder-nsdi25} reports that manually locating a faulty machine took more than half an hour on average and sometimes days.
We use \emph{detection} for deciding that the job is unhealthy and \emph{localization} for identifying the component on which recovery should act.
Localization lies on the critical path from the initial alarm to resumed, useful training.

Synchronous execution makes localization difficult by collapsing rank-local causes into shared symptoms.
If one rank never enters a collective, every rank that did enter waits, and a healthy rank may be the first to report the timeout~\cite{flight-recorder26,mycroft-sosp25}.
A synchronous iteration finishes only after its required peers finish, so its duration reflects the slowest dependency rather than identifying whether a delay came from compute, communication, input, or workload imbalance~\cite{holmes-nsdi25,stragglers-osdi25,greyhound-atc25}.
A reduction such as AllReduce combines tensors contributed by multiple ranks; after one wrong contribution has been incorporated into the result and returned to every rank, the corrupted value no longer identifies its source~\cite{drdna-asplos24,aegis-sdc-osdi26,sdchunter-osdi26}.
Distinct software and hardware faults can consequently produce the same job-wide symptom~\cite{minder-nsdi25,aegis-nsdi25,mycroft-sosp25,flight-recorder26}, and the first reporter may be a victim.
Synchronization therefore leaves multiple causes and candidate ranks consistent with the same alarm, so localization requires additional evidence that distinguishes ranks~\cite{minder-nsdi25,mycroft-sosp25,flight-recorder26}.

Failures are \emph{latent} when synchronization exposes a job-wide symptom without an authoritative signal that identifies the component on which recovery should act.
Although ranks synchronize during training, they remain independent operating-system processes; an out-of-memory kill, uncaught exception, or device error can terminate one rank before its peers~\cite{torchelastic-docs26,dcgm-docs26}.
Such fail-stop failures fall outside this category because the launcher or device-health interface supplies an authoritative rank or device signal.
This paper focuses on three representative latent manifestations.
Silent data corruption (SDC) returns an incorrect numerical value without reporting an error; a recurring straggler continues to make progress but does so abnormally slowly; and a hang leaves all visible participants stopped without naming the rank that initiated the stall.
These manifestations are not exhaustive; we emphasize them because their symptoms do not identify a culprit and each requires a different form of evidence.

Existing mechanisms provide essential detection, observability, and recovery functions, but their evidence does not always identify where to act for the target latent failures.
Launchers identify exited processes~\cite{torchelastic-docs26}, collective watchdogs identify non-completion~\cite{flight-recorder26}, device telemetry reports known hardware errors~\cite{dcgm-docs26}, and checkpoint systems restore previously stored state~\cite{checknrun-nsdi22,gemini-sosp23}.
Tracing, online profiling, metric models, and post-mortem log correlation improve visibility~\cite{megascale-nsdi24,minder-nsdi25,aegis-nsdi25,holmes-nsdi25,eroica-nsdi26,flight-recorder26}, but they do not supply a general localization oracle for latent failures.
Instrumentation may depend on framework internals, monitors inside the training process may become unable to report when that process blocks or terminates, and synchronized logs can preserve the shared symptom rather than its origin.
Generic offline health tests run a different workload and operating regime; production evidence shows that they can miss defects triggered by particular training inputs and execution units~\cite{superbench-tocs26,sdchunter-osdi26}.
SDC also complicates recovery: corruption can perturb parameter updates and persist in model state across steps~\cite{ma-sdc-arxiv25,altenbernd-sdc-arxiv26}, so a byte-consistent checkpoint may already contain corrupted state.
The newest checkpoint may reintroduce corruption, whereas conservative rollback discards healthy work.

These gaps motivate \emph{\system}, a runtime Symmetric Consensus OUTlier detection framework for localizing different failure manifestations.
Its design follows from three observations about synchronous training.
First, failures are rare relative to healthy ranks: an individual failure usually leaves most ranks operating normally~\cite{megascale-nsdi24,minder-nsdi25,aegis-nsdi25}.
Second, practical LLM pre-training jobs typically use data parallelism to create equivalent replicas that execute the same model computation graph at the same training coordinate~\cite{megascale-nsdi24,llama3-arxiv24,fsdp-pvldb23}.
Third, a rank-local failure breaks the behavioral symmetry among these equivalent replicas.
The affected rank may schedule collectives in a different order, take longer to execute, or produce numerical values that differ from healthy ranks~\cite{flight-recorder26}.

These observations suggest a simple design principle: \textbf{localize failures as outliers through strict-majority consensus among equivalent replicas.}
\system groups replicas that share a computation graph and aligns their training coordinate, input, and randomness.
Within each group, majority behavior provides a live, workload-relative reference for progress, elapsed time, or numerical output.
The principle is simple but general: it converts redundancy already present in distributed training into online diagnostic evidence.
It requires neither an absolute health threshold nor a preselected golden rank, exercises the production model rather than a displaced health test, and produces an outlier verdict without offline reconstruction from synchronized logs.

Turning this principle into continuous runtime diagnosis raises three fundamental systems challenges.
First, \system must establish cross-rank comparability before majority voting is meaningful.
Progress is meaningful only when ranks report it in the same training coordinate system, timing is meaningful only for equivalent work, and numerical output is comparable only when benign input and randomness differences are removed.
Second, \system must keep diagnostic evidence available when the training path itself is blocked or suspect.
A hang therefore requires an observer that remains responsive when the trainer blocks, whereas straggler and SDC diagnosis must exercise the live accelerator to preserve production execution conditions.
Third, \system must minimize the system overhead of continuous diagnosis.
Duplicating full iterations or exchanging full tensors would make continuous diagnosis impractical.

\system realizes this principle by making replica evidence comparable, collecting it through failure-appropriate paths, and limiting the work performed for diagnosis.
For comparability, \system aligns training coordinates and inputs, executes equivalent work, and encodes each manifestation in a dedicated evidence schema.
Hang schemas use ordered collective fingerprints, straggler schemas use elapsed time under equivalent work, and SDC schemas use deterministic numerical signatures.
The \emph{Consensus Collective Communication} (C3) abstraction completes this path by evaluating the resulting payloads against the strict majority while preserving which ranks disagree.
For evidence availability and fidelity, \system combines an out-of-band CPU observer with in-situ replay.
An out-of-band CPU process remains available to compare progress after the training group hangs, while in-situ replay exercises recurring stragglers and SDC beside the live job so that model state, kernels, allocations, communication, and operating pressure remain present.
For low overhead, \system combines coverage-based replay scheduling, layer sampling, configurable cadence, rotating coverage, and compact-signature exchange.
Coverage-based scheduling selects representative shapes for recurring execution paths in dynamic MoE workloads instead of continuously replaying every observed shape.
These controls avoid duplicated full iterations and full-tensor exchange while accumulating diagnostic coverage over time.

Together, these mechanisms let \system localize hangs, stragglers, and SDC online.
For a hang, \system reports a rank when its collective sequence or metadata differs from the majority.
For a straggler, \system identifies a rank that remains slower than equivalent peers under controlled replay.
For SDC, \system reports the divergent rank and replay surface whose numerical signature disagrees with the healthy majority.
\system also uses replay verdicts to certify model checkpoints, allowing failure recovery to use in-memory checkpoints when numerically trusted and otherwise fall back to the latest verified checkpoint.
These localization and recovery verdicts are produced online without a separate offline log-analysis step.

\system supports PyTorch~\cite{pytorch-neurips19}, Megatron-Core~\cite{megatron-core-docs26}, TorchTitan~\cite{torchtitan-iclr25}, and DeepSpeed~\cite{zero-sc20} through public interfaces.
It modifies neither training loops nor framework source.
\system also integrates checkpoint certification with GEMINI~\cite{gemini-sosp23}.
These choices make \system an additive service, not a second training runtime.

In summary, this paper makes the following contributions:
\begin{itemize}
  \item We introduce the design principle of identifying outliers through strict-majority consensus among equivalent replicas to localize latent training failures online.
  \item We design C3, an out-of-band hang observer, and in-situ replay to turn this principle into actionable evidence for hangs, stragglers, and SDC.
  \item We use clean replay coverage to certify recovery checkpoints, excluding candidates whose numerical integrity remains unverified.
  \item We implement \system across training frameworks and integrate it with GEMINI.
\end{itemize}

\section{Background}

\subsection{Resilient LLM pre-training}

LLM pre-training iteratively optimizes model parameters over massive token corpora, and a complete run can occupy a cluster for months~\cite{llama3-arxiv24,opt-arxiv22}.
Each iteration computes forward and backward passes, aggregates gradients, and updates the model and optimizer state.
Different parallelism dimensions synchronize through process-group communication: data and tensor parallelism use reduction and gather collectives, context parallelism exchanges attention state, expert parallelism routes tokens with AllToAll, and pipeline parallelism transfers activations and gradients between stages~\cite{pytorch-ddp-pvldb20,fsdp-pvldb23,megatron-arxiv19,megatron-core-docs26,gshard-arxiv20,gpipe-neurips19}.
Because each communication dependency waits for all required participants, an iteration advances no faster than its slowest required rank and stops if any required rank stops participating~\cite{flight-recorder26}.

At training scale, failures are routine: production studies report software exceptions, host and accelerator faults, network failures, and performance degradation~\cite{megascale-nsdi24,minder-nsdi25,aegis-nsdi25,stragglers-osdi25}.
These events fall into two natural categories based on the required recovery action: software failures such as process crashes or input-pipeline exceptions can often be cleared by restarting processes, whereas hardware faults require the affected resource to be replaced, repaired, or excluded~\cite{gemini-sosp23}.
Either class can stall one rank and thereby idle every rank that depends on it.

Resilient pre-training organizes recovery into three interdependent stages: checkpointing, diagnosis, and restart.
\emph{Checkpointing} periodically captures a globally consistent recovery point containing the model, optimizer, and other state required to resume training~\cite{checknrun-nsdi22}.
\emph{Diagnosis} detects that execution is unhealthy and localizes the rank, host, accelerator, or communication path on which recovery should act.
\emph{Restart} applies that action, reconstructs the distributed job, and restores all ranks from the latest checkpoint known to be complete and clean~\cite{torchelastic-docs26}.

High-frequency in-memory checkpointing shifts the recovery bottleneck toward diagnosis.
When a rank stops participating, its peers cannot commit the current iteration, and the job must resume from a mutually consistent model and optimizer state.
Frequent memory-resident checkpoints reduce checkpoint creation and retrieval time while bounding the computation lost to rollback~\cite{gemini-sosp23}.
Diagnosis remains unavoidable: until it identifies a remediation target and a clean recovery point, the accelerator reservation cannot return to useful work.
This paper therefore focuses on reducing this diagnostic interval.

\subsection{Latent failures and their localizations}

Many failures are easy to detect and localize because they produce an explicit component-level error.
For example, an out-of-memory kill or uncaught exception terminates a particular process, a scheduler revocation names a host, and device-health telemetry reports a specific accelerator fault~\cite{torchelastic-docs26,dcgm-docs26}.
The launcher or platform monitor can map such a signal directly to the resource that recovery should restart or replace.

A failure is \emph{latent} when the runtime observes an unhealthy job but receives no authoritative component-level signal for recovery.
Synchronization can propagate a rank-local effect into a shared symptom, making the same underlying fault explicit when reported and latent when only its effect is visible.
This paper focuses on three representative latent manifestations: hangs, stragglers, and silent data corruption (SDC), each of which obscures different information required for recovery.

\noindent \textbf{Hangs obscure which rank stopped useful progress first.}
If one rank blocks before entering a collective, every peer that has entered the collective waits, and a healthy peer may be the first to report a timeout.
A rank that issues a different collective sequence or incompatible metadata can produce the same visible stall as a blocked host thread, device kernel, or communication endpoint~\cite{mycroft-sosp25,flight-recorder26}.
The timeout identifies a collective that did not complete, but it does not by itself identify the initiating rank or distinguish protocol divergence from a failure along an otherwise compatible execution path.

\noindent \textbf{Stragglers obscure which rank is intrinsically slow.}
Because a synchronized dependency completes no faster than its slowest participant, one delayed rank increases the observed iteration time of all ranks that wait for it.
End-to-end iteration time therefore detects lost throughput but cannot separate slow computation or communication from time spent waiting on another rank.
Raw per-rank durations are also ambiguous because legitimate input, routing, and workload differences can change the amount of work assigned to a rank.
Production studies accordingly report heterogeneous causes, including compute degradation, communication delay, input stalls, and workload imbalance~\cite{holmes-nsdi25,stragglers-osdi25,greyhound-atc25}.
Localization requires evidence that a rank remains slower than peers while they perform equivalent work.

\noindent \textbf{SDC obscures both the source of corruption and the safe recovery point.}
An SDC event produces a wrong but valid numerical value without raising an exception or device error.
When a collective incorporates one corrupted contribution, its result can distribute the error to healthy ranks and erase the identity of the source~\cite{drdna-asplos24,aegis-sdc-osdi26,sdchunter-osdi26}.
Subsequent optimizer updates can persist the error in model state, allowing a successfully written checkpoint to contain numerically corrupted parameters~\cite{ma-sdc-arxiv25,altenbernd-sdc-arxiv26}.
Loss curves or activation statistics may reveal that execution is anomalous, but evidence collected after propagation need not identify the rank that introduced the error or the newest checkpoint preceding it.
Safe recovery therefore requires both source localization and positive evidence that the selected checkpoint remains clean.

Production traces quantify both prevalence and cost: stragglers affected 42.5\% of jobs and wasted 10.4\% of allocated GPU-hours, while manual faulty-machine localization took over 30 minutes on average and sometimes days~\cite{stragglers-osdi25,minder-nsdi25}.
For latent failures, localization therefore remains a prerequisite for recovery because no component-level signal supplies an action.
Checkpoint and restart mechanisms cannot select a resource or clean recovery point until diagnostic evidence narrows the culprit.
Hangs, stragglers, and SDC require rank-distinguishing evidence tailored to the information that each symptom obscures.

\section{\system Overview}

This section reframes latent failures as behavioral outliers and derives \system from that perspective.
It validates the conditions required for consensus detection and then presents \system's design principle and architecture.

\subsection{Latent Failures Are Outliers}

Existing diagnosis systems demonstrate the value of relative evidence for distributed training.
Minder~\cite{minder-nsdi25} identifies failure-related metric patterns, Holmes~\cite{holmes-nsdi25} compares irregularities across workers, and EROICA~\cite{eroica-nsdi26} correlates differential observations across the stack.
Their localization accuracy, however, depends on a predefined set of telemetry signals and fault-specific analysis rules.
A fault is attributable only when it creates a rank-specific pattern in exported metrics, logs, or traces; when synchronization makes healthy and faulty ranks expose the same stall or slowdown, these systems may observe the job-wide symptom without identifying its source.

\system instead defines the localization target independently of any particular telemetry source.
A latent failure makes a rank a behavioral outlier when its progress, execution time, or numerical result departs from what it would have produced without the fault under the same work and operating conditions~\cite{mycroft-sosp25,holmes-nsdi25,aegis-sdc-osdi26}, as listed in~Table~\ref{tab:failure-outliers}.
Because this counterfactual is unobservable, localization requires an observable reference with the same expected behavior as the rank being evaluated.

Such a reference can be temporal, using the same rank's history, or spatial, using concurrent ranks.
Temporal comparison can in principle expose the three latent failure manifestations by detecting a stalled progress signal, a latency increase, or a changed result when the same computation recurs.
In LLM training, however, expected progress, latency, and numerical output vary across layers, inputs, model states, and MoE routing decisions.
Consequently, a temporal detector must model these phases and set phase-specific decision boundaries to avoid confusing legitimate variation with a failure.
Temporal history is therefore useful for corroborating trends and shared slowdowns, but it is an unstable primary reference for a phase-varying workload.

\system instead uses spatial comparison as its primary localization mechanism.
Concurrent peers performing equivalent work provide the reference; when a strict majority agrees, their result estimates the evaluated rank's fault-free behavior.
Because the comparison is concurrent, it does not rely on measurements collected under an earlier model state, input regime, or cluster load.
Spatial comparison is valid only among ranks with the same expected behavior.
For progress comparison, peers occupy the same tensor-parallel shard, pipeline stage, context-parallel shard, and expert assignment while belonging to different data-parallel or FSDP instances.
For timing and numerical replay, \system additionally aligns model state, input, randomness, and execution conditions.

\begin{table}[t]
\centering
\small
\renewcommand{\arraystretch}{1.08}
\begin{tabular}{@{}p{0.20\columnwidth}p{0.73\columnwidth}@{}}
\hline
Manifestation & Spatial behavioral outlier \\
\hline
Hang & One rank reports different progress or collective intent after the group stalls. \\
Straggler & One rank is persistently slower than its peers on controlled equivalent work. \\
SDC & One rank produces a different deterministic value for the same computation. \\
\hline
\end{tabular}
\caption{Spatial behavioral outliers for latent failures.}
\label{tab:failure-outliers}
\end{table}

\system's spatial comparison requires three conditions: a failure creates a rank-local divergence in a compared behavior, faulty observations remain a minority within a peer group, and training provides equivalent peers.
Table~\ref{tab:failure-outliers} establishes the first condition: hangs, stragglers, and SDC create progress, timing, and numerical outliers, respectively.

Production reports indicate that the faulty ranks implicated by an individual incident typically form a small minority, even when synchronization spreads the symptom across the job.
ByteRobust~\cite{byterobust-sosp25} reports that large-scale training failures normally occur independently on individual nodes, that simultaneous multi-node failures are extremely rare, and that only one or two nodes are typically faulty in a 9,600-GPU job.
MegaScale~\cite{megascale-nsdi24} likewise documents how one faulty GPU worker can cascade into a cluster-wide NCCL stall.
\system encodes this operational pattern as a strict-majority requirement instead of assuming exactly one faulty rank.

Hybrid parallelism routinely leaves equivalent peers in both dense and MoE pre-training, as listed in~Table~\ref{tab:replica-evidence}.
Tensor, pipeline, context, and expert parallelism partition model work, whereas data parallelism repeats the resulting partitions.
The non-data-parallel coordinates of a rank are its tensor shard, pipeline stage, context shard, and assigned expert set.
Ranks that match at each of these positions but differ along the data-parallel or FSDP dimension execute corresponding model partitions and therefore form candidate peer groups.
Recent frontier MoE systems reinforce this pattern: Kimi K2.5~\cite{kimi-k25-arxiv26} uses ZeRO-1 data parallelism, GLM-5~\cite{glm5-arxiv26} shards gradients across data-parallel ranks, and DeepSeek-V4~\cite{deepseekv4-arxiv26} synchronizes MoE gradients across data-parallel ranks, although their reports do not disclose a fixed pre-training DP degree.

\begin{table}[t]
\centering
\small
\renewcommand{\arraystretch}{1.08}
\begin{tabular}{@{}p{0.30\columnwidth}r p{0.27\columnwidth}r@{}}
\hline
Model & GPUs & Parallelism & Peer degree \\
\hline
Megatron 1T~\cite{megatron-sc21} & 3,072 & TP8, PP64, DP6 & 6 \\
MegaScale 175B~\cite{megascale-nsdi24} & 12,288 & TP8, PP8, DP192 & 192 \\
GLM-130B~\cite{glm130b-iclr23} & 768 & TP4, PP8, DP24 & 24 \\
Llama 3 405B~\cite{llama3-arxiv24} & 16,384 & TP8, PP16, DP128 & 128 \\
Skywork-MoE~\cite{skyworkmoe-arxiv24} & 1,536 & PP12, EDP4, DP32 & 32 \\
MegaScale-MoE~\cite{megascale-moe-eurosys26} & 1,440 & PP15, EP8, DP12 & 12 \\
\hline
\end{tabular}
\caption{Published dense and MoE configurations with equivalent peer groups.}
\label{tab:replica-evidence}
\end{table}

\subsection{Architecture Overview}

\begin{figure*}[t]
\centering
\begin{tikzpicture}[
  font=\sffamily\footnotesize,
  stack/.style={draw=black!65, rounded corners=1.2pt, fill=white, minimum width=3.25cm, minimum height=0.62cm, align=center},
  band/.style={draw=black!35, rounded corners=2pt, fill=white},
  layername/.style={font=\sffamily\normalsize\bfseries, anchor=center},
  adapter/.style={draw=black!65, rounded corners=1.2pt, fill=white, minimum width=14.7cm, minimum height=0.55cm, align=center},
  topology/.style={draw=black!65, rounded corners=1.2pt, fill=white, minimum width=11.8cm, minimum height=0.55cm, align=center},
  component/.style={draw=black!65, rounded corners=1.2pt, fill=white, minimum height=0.92cm, text width=5.4cm, align=center},
  decision/.style={draw=black!65, rounded corners=1.2pt, fill=white, minimum height=0.82cm, text width=5.0cm, align=center},
  output/.style={draw=black!65, rounded corners=1.2pt, fill=white, minimum height=0.72cm, text width=5.9cm, align=center},
  flow/.style={-{Latex[length=2.0mm,width=1.3mm]}, draw=black!75, line width=0.55pt}
]
\node[stack] (pytorch) at (-5.7,0) {PyTorch};
\node[stack] (torchtitan) at (-1.9,0) {TorchTitan};
\node[stack] (megatron) at (1.9,0) {Megatron-Core};
\node[stack] (deepspeed) at (5.7,0) {DeepSpeed};

\draw[band] (-7.7,-0.72) rectangle (7.7,-2.66);
\node[layername] at (0,-0.98) {Integration layer};
\node[adapter] (adapter) at (0,-1.45) {\textbf{Framework adapters}\quad public module, optimizer, process-group, and checkpoint interfaces};
\node[topology] (topology) at (0,-2.35) {\textbf{Topology manager (Sec.~\ref{sec:peer-groups})}\quad parallel coordinates $\rightarrow$ equivalent peer groups};

\draw[band] (-7.7,-2.86) rectangle (7.7,-4.58);
\node[layername] at (0,-3.12) {Evidence layer};
\node[component] (replay) at (-3.55,-3.87) {\textbf{In-situ layer replay (Sec.~\ref{sec:replay})}\\captured layer $\rightarrow$ live replay\\numerical signatures and execution time};
\node[component] (oob) at (3.55,-3.87) {\textbf{Trainer-independent OOB observer (Sec.~\ref{sec:oob-hangs})}\\shared progress $\rightarrow$ CPU comparison\\progress and collective fingerprints};

\draw[band] (-7.7,-4.78) rectangle (7.7,-8.72);
\node[layername] at (0,-5.04) {Decision layer};
\node[decision, text width=6.6cm] (c3) at (0,-5.72) {\textbf{Consensus Collective Communication (Sec.~\ref{sec:c3})}\\exact majority and robust timing outliers};
\node[decision, text width=6.6cm] (record) at (0,-6.98) {\textbf{Evidence record (Sec.~\ref{sec:c3})}\\healthy, rank outlier, or group stall};
\node[output] (gate) at (-3.20,-8.25) {\textbf{Checkpoint gate (Sec.~\ref{sec:checkpoint-certification})}\\clean numerical replay required};
\node[output] (policy) at (3.20,-8.25) {\textbf{External recovery policy}\\continue, restart, quarantine, or diagnose};

\foreach \framework in {pytorch,torchtitan,megatron,deepspeed}
  \draw[flow] (\framework.south) -- (\framework.south |- adapter.north);
\draw[flow] (adapter.south) -- (topology.north);
\draw[flow] (replay.north |- topology.south) -- (replay.north);
\draw[flow] (oob.north |- topology.south) -- (oob.north);
\draw[flow] (replay.south) -- ([xshift=-2.5cm]c3.north);
\draw[flow] (oob.south) -- ([xshift=2.5cm]c3.north);
\draw[flow] (c3.south) -- (record.north);
\draw[flow] ([xshift=-2.5cm]record.south) -- (gate.north);
\draw[flow] ([xshift=2.5cm]record.south) -- (policy.north);
\end{tikzpicture}
\caption{\system separates runtime failure localization into three layers.
The integration layer establishes equivalent peers across frameworks, the evidence layer preserves numerical, timing, and hang observations, and the decision layer turns peer comparisons into localization and checkpoint decisions.}
\Description{PyTorch, TorchTitan, Megatron-Core, and DeepSpeed connect to an integration layer containing framework adapters and a topology manager.
Vertical arrows connect the topology manager to in-situ layer replay and a trainer-independent out-of-band observer in the evidence layer.
Both observation mechanisms feed Consensus Collective Communication in the decision layer, which produces a healthy result, rank outlier, or group stall for the checkpoint gate and an external recovery policy.}
\label{fig:architecture}
\end{figure*}
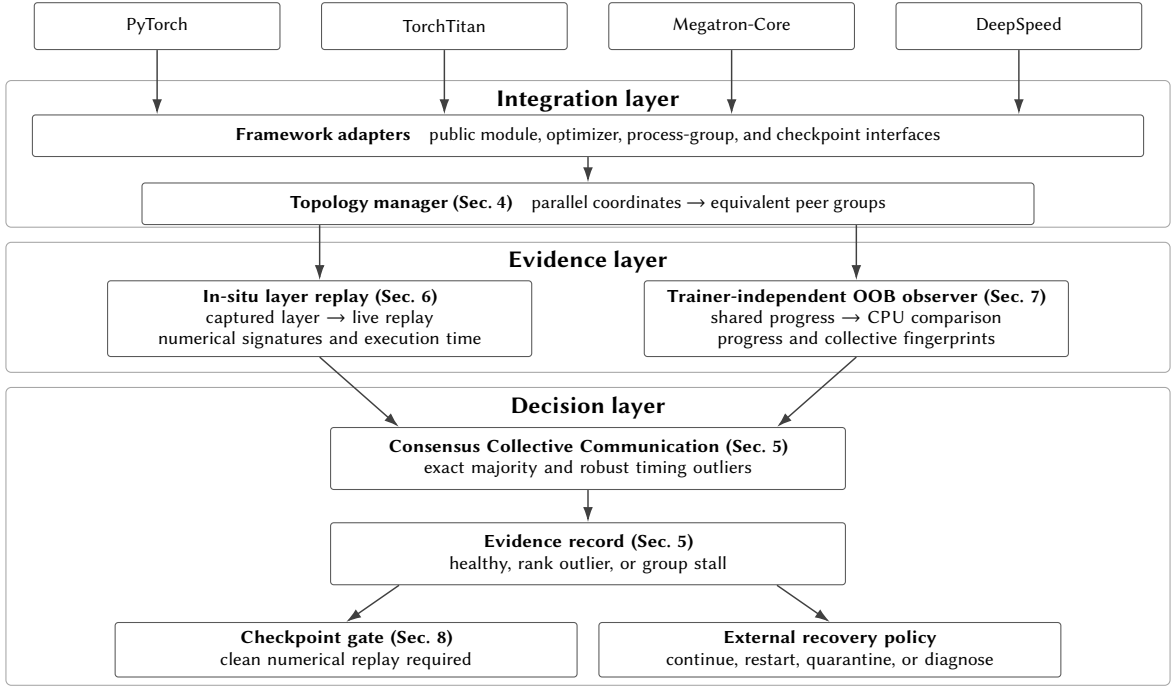

The behavioral-outlier formulation yields one design principle: localize failures as outliers through strict-majority consensus among equivalent replicas.
Within an aligned peer group, the majority supplies a live reference for expected progress, execution time, or deterministic numerical output, and minority behavior identifies localization candidates.
\system needs neither an absolute health threshold nor a preselected golden rank.

\system organizes runtime diagnosis into three layers, as shown in~Figure~\ref{fig:architecture}.
The integration layer connects \system to each framework and determines which ranks perform equivalent work.
The evidence layer keeps failure-specific observations available under the conditions that produce them.
The decision layer compares them and produces localization and checkpoint decisions.

\noindent \textbf{Integration layer.}
Strict-majority consensus is meaningful only among ranks expected to perform equivalent work, so this layer normalizes framework structure and constructs peer groups.
A framework adapter uses public module-hook, optimizer, process-group, and checkpoint interfaces to expose the model architecture, communication groups, optimizer boundary, and checkpoint boundary.
The topology manager combines this structure with each rank's parallel coordinates to group ranks that execute the same model partition at the same training coordinate (Section~\ref{sec:peer-groups}).
The application continues to issue its ordinary forward, backward, collective, and optimizer operations without framework-source changes or diagnostic branches in the training loop.

\noindent \textbf{Evidence layer.}
\system uses two observation mechanisms because SDC and straggler diagnosis must preserve live accelerator conditions, whereas hang diagnosis must survive a blocked trainer.
\system diagnoses SDC and stragglers through bounded in-situ replay on the live accelerator (Section~\ref{sec:replay}).
Replay preserves the training job's model state, kernels, allocations, communication activity, thermal conditions, and compute and memory pressure while yielding compact numerical signatures and execution times for peer comparison.
A hang may block both the trainer and its NCCL communicator, so neither can reliably participate in diagnosis.
\system handles hangs with an out-of-band (OOB) CPU observer that reads progress and pending-collective fingerprints published by training hooks through shared memory (Section~\ref{sec:oob-hangs}).
The observer remains responsive on a control communicator separate from the blocked training communicator and compares peers after a progress timeout.

\noindent \textbf{Decision layer.}
The decision layer distinguishes healthy execution, localized rank divergence, and group-scoped stalls.
\system's \emph{Consensus Collective Communication} (C3) applies exact majority comparison to progress, collective fingerprints, and numerical signatures, and robust statistical comparison to replay timing (Section~\ref{sec:c3}).
C3 returns \emph{healthy} when scheduled comparisons agree and an evidence record naming the disagreeing ranks when a strict majority identifies an outlier.
For hangs, divergent progress or collective fingerprints expose a visible software or control-flow mismatch, whereas agreement after a progress timeout classifies the incident as an equal-progress runtime or transport stall for further hardware or fabric diagnosis.
An external recovery policy takes no action on a healthy result, restarts the job in place for a localized software or control-flow failure, and replaces or quarantines a resource for a confirmed hardware failure.
An SDC that reaches a parameter or optimizer update becomes part of subsequent training state, so a newly captured checkpoint can preserve the corruption and reintroduce it after restart.
\system also uses replay evidence to decide whether recovery can use the latest checkpoint or must fall back to a verified checkpoint (Section~\ref{sec:checkpoint-certification}).

\section{Forming Equivalent Peer Groups}
\label{sec:peer-groups}

\system forms equivalent peer groups by mapping framework-specific parallelism into logical mesh addresses.
Peers retain the same model-parallel role while differing only along an eligible data-replica or state-shard dimension.

LLM training maps ranks onto replica, state-sharding, tensor-, pipeline-, context-, and expert-parallel dimensions~\cite{pytorch-ddp-pvldb20,fsdp-pvldb23,megatron-arxiv19,gpipe-neurips19,gshard-arxiv20,llama3-arxiv24}.
Let $\mathcal{R}$ be the set of training ranks, and represent each rank by the logical parallelism-mesh address
\[
r=(d_r,s_r,t_r,p_r,c_r,e_r).
\]
Here, $d_r$ is the data-replica coordinate; $s_r$ is the FSDP state-shard coordinate; and $t_r$, $p_r$, $c_r$, and $e_r$ are the tensor-parallel position, pipeline stage, context-parallel position, and expert partition, respectively.
Ordinary data parallelism varies $d$ and has a degree-one $s$ dimension, whereas pure FSDP varies $s$ and has a degree-one $d$ dimension.
Any other dimension absent from a configuration also has degree one and imposes no constraint.
The address is logical rather than framework-native: \system reads it from a PyTorch device mesh~\cite{torchtitan-iclr25} or derives it from the process groups configured by Megatron-Core~\cite{megatron-core-docs26} and DeepSpeed~\cite{zero-sc20}.

Ranks can be compared only when the coordinates that determine their assigned computation match.
Matching $t$ selects the same tensor shard, matching $p$ selects the same layers, and matching $e$ selects the same expert partition.
Matching $c$ is also necessary because context-parallel ranks hold different sequence shards and participate in position-specific attention communication even though they replicate model parameters.
\system therefore compares only ranks at the same context-parallel position.

With these model-parallel coordinates fixed, group formation has two cases: \system varies $d$ when natural replicas exist and varies $s$ when state sharding is the only source of peers.

\noindent \textbf{Natural replica peers.}\hspace{0.35em}
For a rank $r$ in a job whose data-replica dimension has degree greater than one, \system forms
\[
\begin{aligned}
G_{\mathrm{rep}}(r)=\{q\in\mathcal{R}\mid {}&s_q=s_r,\ t_q=t_r,\ p_q=p_r,\\
&c_q=c_r,\ e_q=e_r\}.
\end{aligned}
\]
The unconstrained $d_q$ coordinate selects corresponding ranks across data-parallel replicas while every state-shard and model-parallel coordinate remains fixed.

For example, consider eight global ranks arranged as a four-way data-parallel by two-way tensor-parallel mesh with logical shape $(d,s,t,p,c,e)=(4,1,2,1,1,1)$.
Under row-major placement, the rank at logical position $(d,t)$ has global rank $2d+t$.
Therefore, the two equivalent peer groups are $G_{\mathrm{rep}}(0)=\{0,2,4,6\}$ and $G_{\mathrm{rep}}(1)=\{1,3,5,7\}$.

\noindent \textbf{State-shard peers.}\hspace{0.35em}
When the data-replica dimension has degree one, no natural replica group exists, so \system instead forms peers along the FSDP state-shard dimension.
FSDP ranks can form an equivalent peer group because they execute the same forward and backward computation after the standard parameter AllGather~\cite{fsdp-pvldb23}.
\begin{samepage}
For pure FSDP, \system forms
\[
\begin{aligned}
G_{\mathrm{shard}}(r)=\{q\in\mathcal{R}\mid {}&t_q=t_r,\ p_q=p_r,\\
&c_q=c_r,\ e_q=e_r\}.
\end{aligned}
\]
\end{samepage}
The unconstrained $s_q$ coordinate selects the ranks that shard the same model-parallel partition.

DP and FSDP expose the same computation graph but differ in what each rank holds before an operation begins.
DP ranks already hold corresponding model state, whereas FSDP ranks hold different parameter and optimizer-state shards.
\system bridges this difference by synchronizing the captured input and the state used by the selected diagnostic operation, so every state-shard peer executes that operation with the same values.

\noindent \textbf{Peer-group size.}\hspace{0.35em}
Peer-group size determines how precisely \system can act on a disagreement.
A singleton group provides no comparison.
A two-rank group can detect disagreement but cannot identify the faulty rank.
\system therefore returns a group-scoped result, allowing recovery to rule out both ranks.
A group of at least three can identify minority outliers whenever healthy observations form a strict majority.
For a group of size $N$, rank attribution therefore requires more than $N/2$ healthy members.
\section{Consensus Collective Communication}
\label{sec:c3}


Existing collectives exchange rank-local values, but they do not decide which ranks disagree with their peers.
\system instead uses \emph{Consensus Collective Communication} (C3), a diagnostic collective abstraction that combines a standard AllGather with deterministic consensus to return the disagreeing ranks.

\subsection{Evidence and verdict interface}

C3 maps one diagnostic object from each rank of an equivalent peer group to a result $R=(s,B,E)$.
Let $G=(r_0,\ldots,r_{N-1})$ be an ordered peer group, and let $x_i$ be any supported diagnostic object supplied by $r_i$.
C3 first converts $x_i$ to a comparable evidence $e_i$ (\S~\ref{sec:c3_comparable_evidence}) and an AllGather gives every participant $E=(e_0,\ldots,e_{N-1})$ before each rank applies the same deterministic comparison rule locally.
The result contains a status $s$, an $N$-bit outlier bitmap $B$, and the gathered evidence $E$; $B[i]=1$ attributes the divergence to peer $r_i$.
This decentralized evaluation avoids a coordinator and ensures that all responsive peers derive the same result.

C3 distinguishes three statuses in $s$ so that an all-zero bitmap is not ambiguous.
\emph{Agree} means that the evidence satisfies the selected agreement rule and $B=0_N$.
\emph{Attributed} means that the rule identifies one or more outliers in $B$.
\emph{Inconclusive} means that peers disagree but the evidence does not justify rank attribution, such as an exact comparison without a strict majority; C3 returns $B=0_N$ rather than naming an arbitrary culprit.
\system stores the result with the ordered peer-to-global-rank mapping so its failure handler can map bitmap positions to training ranks.

C3 deliberately separates comparison from interpretation.
The same interface accepts progress coordinates, collective fingerprints, tensors or structured tensor outputs, and execution times.
Algorithm~\ref{alg:c3} lists the common protocol and makes the returned result explicit.

\begin{algorithm}[t]
\caption{Consensus Collective Communication}
\label{alg:c3}
\KwIn{Diagnostic object $x_i$ from rank $r_i$; comparison mode $\mathit{mode}\in\{\textsc{Exact},\textsc{Statistical}\}$; ordered peer group $G=(r_0,\ldots,r_{N-1})$}
\KwOut{Result $R=(s,B,E)$ containing a status, an outlier bitmap, and the gathered evidence}
\Fn{\textsc{C3}$(x_i,\mathit{mode},G)$}{
  {$e_i \gets$ produce comparable evidence from  $x_i$\;}
  {$E \gets \operatorname{AllGather}(G,e_i)$\;}
  \eIf{$\mathit{mode}=\textsc{Exact}$}{
    {$(e^*,c^*) \gets$ most frequent value in $E$ and its count\;}
    \If{$c^* \le |G|/2$}{
      {\Return $(\textsc{Inconclusive},0_N,E)$\;}
    }
    \ForEach{$r_j\in G$}{
      {$B[j] \gets \mathbf{1}\{e_j \ne e^*\}$\;}
    }
  }{
    {$(m,d) \gets (\operatorname{median}(E),\operatorname{RobustScale}(E))$\;}
    \If{$d=0$}{
      {\Return $(\textsc{Agree},0_N,E)$\;}
    }
    \tcp{$\kappa>0$ is the sensitivity multiplier}
    \ForEach{$r_j\in G$}{
      {$B[j] \gets \mathbf{1}\{|e_j-m|>\kappa d\}$\;}
    }
  }
  \If{$\exists j:B[j]=1$}{
    {\Return $(\textsc{Attributed},B,E)$\;}
  }
  {\Return $(\textsc{Agree},0_N,E)$\;}
}
\end{algorithm}

\subsection{Producing comparable evidence}
\label{sec:c3_comparable_evidence}

C3 can take scalars or tensors as its payload.
Small fixed-size values, including progress coordinates, execution times, and collective fingerprints, are supplied as comparable evidence directly.
Gathering a large GPU-resident tensor, however, would make diagnostic traffic scale with model-state or activation size.
In addition, element-wise comparison among multiple tensors on each GPU can incur costly overheads.
For each checked tensor $T_i$---such as a layer output, input gradient, parameter gradient, or updated weight---rank $r_i$ therefore supplies
\[
e_i=H_w(\operatorname{dtype}(T_i),\operatorname{shape}(T_i),\operatorname{bytes}(T_i)),
\]
where $H_w$ is a deterministic, position-sensitive hash that produces a $w$-bit signature over the tensor metadata and bytes.

C3 is independent of signature width, while \system's current implementation uses 64-bit signatures.
\system performs the bulk signature fold on the tensor's accelerator and transfers only compact intermediate state for finalization, avoiding a full device-to-host copy or a full-tensor exchange.
Once each rank has produced comparable evidence, C3 uses exact consensus for deterministic values and statistical consensus for timing values.

\subsection{Exact consensus}

Exact consensus compares evidence whose healthy values should be identical.
Progress coordinates, collective fingerprints, and deterministic numerical signatures all satisfy this requirement.
Algorithm~\ref{alg:c3} lists the exact-consensus procedure.
\system first selects the most frequent value $v^*$ and its count $c^*$ (Line 4).
If $c^*>N/2$, \system marks every rank whose evidence differs from $v^*$ (Lines 7--8).
The common return logic reports \emph{attributed} when at least one bit is set or \emph{agree} when none is set (Lines 15--17).
If $c^*\le N/2$, \system returns \emph{inconclusive} with a zero bitmap (Lines 5--6) because the evidence does not identify a strict-majority reference.
This rule handles several distinct minority values without assuming exactly one faulty rank, while refusing to choose a side in a two-rank disagreement.

Exact consensus is a sparse-fault attribution rule, not Byzantine agreement~\cite{lamport-toplas82} or a correctness oracle.
It assumes that responsive ranks report one value each and that the healthy value holds a strict majority; it does not establish agreement under arbitrary behavior.
A faulty value held by a strict majority is indistinguishable from the expected value, and common-mode corruption that makes every peer agree remains invisible.
Section~\ref{sec:peer-groups} therefore makes the healthy-majority assumption and the available attribution scope explicit before C3 runs.

\subsection{Statistical consensus}

Healthy peers can report slightly different execution times, so C3 uses robust statistical consensus for timing evidence.
For peer timings $T=(\tau_0,\ldots,\tau_{N-1})$, C3 computes their median $m$ and a robust scale $d$, then marks
\[
B[i]=\mathbf{1}\{|\tau_i-m|>\kappa d\}.
\]
The robust scale $d=\operatorname{RobustScale}(T)$ estimates ordinary peer-to-peer timing dispersion without being dominated by a small number of extreme observations.
The sensitivity multiplier $\kappa$ specifies how many units of that dispersion a timing may deviate from the median before C3 marks it as an outlier; a larger $\kappa$ requires a more extreme deviation~\cite{iglewicz-hoaglin93}.
\system first estimates $d$ from the median absolute deviation (MAD), $\operatorname{median}_i|\tau_i-m|$, and falls back to the interquartile range and then a scaled observed range when an estimate is zero.
Consequently, $\operatorname{RobustScale}(T)$ returns zero only when every timing is identical; Lines 11--12 then report \emph{agree} because there is no timing divergence to attribute.

The median remains a useful spatial reference only while degraded observations remain a minority.
Statistical C3 intentionally misses a fleet-wide slowdown because no peer differs from the others, and deviations within $\kappa d$ remain unreported.
Temporal baselines can detect shared or gradual shifts, but without a concurrent spatial differential they do not add rank attribution.

\section{In-Situ Training Replay}
\label{sec:replay}

A pre-training iteration proceeds through data loading, forward pass, backward pass, and the optimizer update~\cite{torchtitan-iclr25}.
Periodically, a selected iteration adds checkpoint I/O after the optimizer update~\cite{gemini-sosp23,checknrun-nsdi22}.
\system observes data-loading and checkpoint-I/O delays at phase boundaries, but replays selected work within other phases because progress alone cannot distinguish ranks.

Data loading and checkpoint I/O expose explicit entry and exit boundaries, so \system localizes their hangs and stragglers from progress reported by the out-of-band (OOB) observers in Section~\ref{sec:oob-hangs}.
At the configured sampling interval, the framework adapter marks when a rank starts and finishes fetching a training batch.
On checkpoint iterations, checkpoint hooks similarly mark the start and completion of checkpoint reads and writes.
If one rank remains inside either phase after peers advance, C3 localizes that rank as an outlier.


\subsection{Dense-model replay}

Progress observation cannot reveal an incorrect result or distinguish slow local work from waiting at synchronization.
\system therefore replays selected forward, backward, and optimizer work to produce rank-distinguishing numerical and timing evidence.

Fixed local tensor shapes make dense-model replay reusable and bounded.
Under a fixed training configuration, tensor shapes remain constant even as their values change.
\system builds recipes for the embedding layer, hidden layers, the language-model output layer, and the optimizer update.
The dense Transformer models we target stack hidden layers with the same module architecture and local tensor shapes~\cite{opt-arxiv22,llama3-arxiv24}, allowing \system to reuse one replay recipe across them.
Each recipe's replay interval can be configured independently.
When a recipe is due, \system executes it between iterations so that replay does not affect the training result.

The embedding and language-model output recipes broadcast captured token indices or a captured hidden-state tensor plus the incoming output gradient, respectively, then compare output and gradient signatures.
The framework adapter preserves vocabulary and tensor-parallel partitions, so C3 compares only peers with the same model-parallel coordinates.

\subsubsection{Hidden-layer replay}

\begin{algorithm}[!t]
\caption{Ordinary FSDP forward and \system replay}
\label{alg:dense-replay}
\KwIn{Module $M$ with shard $\theta_r$; invocation $x_r$; shard group $S_r$; pre-forward RNG state $\rho_r$; equivalent peers $G$}
\Fn{\textsc{FSDP}$(M,x_r,S_r)$}{
  $\theta \gets \operatorname{AllGatherParameters}(S_r,\theta_r)$\;
  $y_r \gets M(x_r, \theta)$\;
  \Return $y_r$\;
}
\Fn{\textsc{Replay}$(M,x_r,S_r,\rho_r,G)$}{
  $(x,\rho) \gets \operatorname{BroadcastSource}(G,x_r,\rho_r)$\;
  $R_{\mathrm{broadcast}} \gets \textsc{C3}((x, \rho),\textsc{Exact},G)$\;
  $(\theta,t_{\mathrm{gather}}) \gets \operatorname{TimedAllGather}(S_r,\theta_r)$\;
  $(y,t_M) \gets \operatorname{TimedCall}(M,\operatorname{Copy}(x, \theta))$\;
  $R_{\mathrm{coll}} \gets \textsc{C3}(\theta,\textsc{Exact},G)$\;
  $R_{\mathrm{sdc}} \gets \textsc{C3}(y,\textsc{Exact},G)$\;
  $R_{\mathrm{gather}} \gets \textsc{C3}(t_{\mathrm{gather}},\textsc{Statistical},G)$\;
  $R_{\mathrm{module}} \gets \textsc{C3}(t_M,\textsc{Statistical},G)$\;
  \Return $(R_{\mathrm{broadcast}}, R_{\mathrm{coll}},R_{\mathrm{sdc}},R_{\mathrm{gather}},R_{\mathrm{module}})$\;
}
\end{algorithm}

Algorithm~\ref{alg:dense-replay} uses an FSDP-sharded hidden layer to contrast ordinary forward execution with replay.
Let $S_r$ denote the FSDP shard group that materializes rank $r$'s parameters, and let $G$ denote the equivalent peer group whose evidence C3 compares.
\textsc{Replay()} preserves normal FSDP parameter materialization but broadcasts the source input and RNG state, computes on copied state, and compares the materialized parameters and module output outside the training result.
Two techniques make replay evidence comparable, while input variation broadens coverage.

\begin{figure*}[!t]
\centering
\begin{tikzpicture}[
  font=\sffamily\footnotesize,
  panel/.style={draw=black!28, rounded corners=2pt, fill=black!1},
  gpu/.style={draw=black!65, rounded corners=1.2pt, fill=white, minimum width=1.38cm, minimum height=0.58cm, align=center, inner sep=1.5pt},
  compute/.style={draw=revisionblue, rounded corners=1.2pt, fill=revisionblue!13, line width=1.0pt, minimum width=1.38cm, minimum height=0.58cm, align=center, inner sep=1.5pt},
  collective/.style={draw=orange!80!black, rounded corners=1.2pt, fill=orange!18, line width=1.0pt, minimum width=1.38cm, minimum height=0.58cm, align=center, inner sep=1.5pt},
  fsdp/.style={draw=black!58, line width=1.4pt},
  slowfsdp/.style={draw=orange!80!black, line width=2.4pt},
  dppeer/.style={draw=revisionblue!72, line width=0.65pt},
  axis/.style={-{Latex[length=1.7mm,width=1.2mm]}, draw=black!72, line width=0.55pt},
  callout/.style={font=\sffamily\scriptsize, align=center, text width=4.55cm}
]

\draw[panel] (-8.0,0.95) rectangle (-2.84,-3.58);
\draw[panel] (-2.54,0.95) rectangle (2.54,-3.58);
\draw[panel] (2.84,0.95) rectangle (8.0,-3.58);
\node[font=\sffamily\bfseries] at (-5.38,0.68) {(a) Logical $4\!\times\!2$ mesh};
\node[font=\sffamily\bfseries] at (0,0.68) {(b) Module-time outlier};
\node[font=\sffamily\bfseries] at (5.38,0.68) {(c) AllGather-time outlier};

\draw[dppeer] (-6.30,0.08) -- (-6.30,-2.32);
\draw[dppeer] (-4.55,0.08) -- (-4.55,-2.32);
\foreach \d/\ra/\rb/\yy in {0/0/1/0.08,1/2/3/-0.72,2/4/5/-1.52,3/6/7/-2.32} {
  \draw[fsdp] (-6.30,\yy) -- (-4.55,\yy);
  \node[gpu] at (-6.30,\yy) {$r_{\ra}$\\[-1pt]{\scriptsize $(d_{\d},s_0)$}};
  \node[gpu] at (-4.55,\yy) {$r_{\rb}$\\[-1pt]{\scriptsize $(d_{\d},s_1)$}};
  \node[font=\sffamily\scriptsize, anchor=west] at (-3.43,\yy) {$S_{\d}$};
}
\draw[axis] (-6.30,-2.79) -- (-4.55,-2.79);
\node[font=\sffamily\scriptsize, anchor=north] at (-5.42,-2.83) {FSDP dimension $s$};
\draw[axis] (-7.35,0.08) -- (-7.35,-2.32);
\node[font=\sffamily\scriptsize, rotate=90, anchor=south] at (-7.40,-1.12) {DP dimension $d$};

\draw[dppeer] (-0.92,0.08) -- (-0.92,-2.32);
\draw[dppeer] (0.92,0.08) -- (0.92,-2.32);
\foreach \d/\ra/\rb/\yy in {0/0/1/0.08,1/2/3/-0.72,2/4/5/-1.52,3/6/7/-2.32} {
  \draw[fsdp] (-0.92,\yy) -- (0.92,\yy);
  \ifnum\ra=4
    \node[compute] at (-0.92,\yy) {$r_{\ra}$\\[-1pt]{\scriptsize high $t_M$}};
  \else
    \node[gpu] at (-0.92,\yy) {$r_{\ra}$};
  \fi
  \node[gpu] at (0.92,\yy) {$r_{\rb}$};
}
\node[callout] at (0,-3.07) {Only $r_4$ differs across DP-equivalent peers\\$\Rightarrow$ computation straggler on $r_4$};

\draw[dppeer] (4.46,0.08) -- (4.46,-2.32);
\draw[dppeer] (6.30,0.08) -- (6.30,-2.32);
\foreach \d/\ra/\rb/\yy in {0/0/1/0.08,1/2/3/-0.72,2/4/5/-1.52,3/6/7/-2.32} {
  \ifnum\d=2
    \draw[slowfsdp] (4.46,\yy) -- (6.30,\yy);
    \node[collective] at (4.46,\yy) {$r_{\ra}$\\[-1pt]{\scriptsize high $t_{\rm gather}$}};
    \node[collective] at (6.30,\yy) {$r_{\rb}$\\[-1pt]{\scriptsize high $t_{\rm gather}$}};
    \node[font=\sffamily\scriptsize\bfseries, anchor=west, text=orange!70!black] at (7.38,\yy) {$S_2$};
  \else
    \draw[fsdp] (4.46,\yy) -- (6.30,\yy);
    \node[gpu] at (4.46,\yy) {$r_{\ra}$};
    \node[gpu] at (6.30,\yy) {$r_{\rb}$};
  \fi
}
\node[callout] at (5.38,-3.07) {Both members of $S_2$ report slow AllGather\\$\Rightarrow$ affected collective group $S_2$, not an endpoint};
\end{tikzpicture}
\caption{Separate module and communication timers identify a computation straggler directly and localize a communication straggler to an affected PG.}
\Description{Three panels show eight GPUs arranged as four data-parallel rows and two FSDP-shard columns.
The first panel labels the DP and FSDP dimensions, with horizontal links denoting four two-rank FSDP AllGather groups and vertical solid lines denoting DP-equivalent peers.
The second panel highlights only rank r4 as having high module time, which indicates a rank-local computation straggler.
The third panel highlights ranks r4 and r5 and their horizontal link as having high AllGather time, which identifies FSDP group S2 as affected but does not identify a faulty endpoint.}
\label{fig:compute-communication-separation}
\end{figure*}
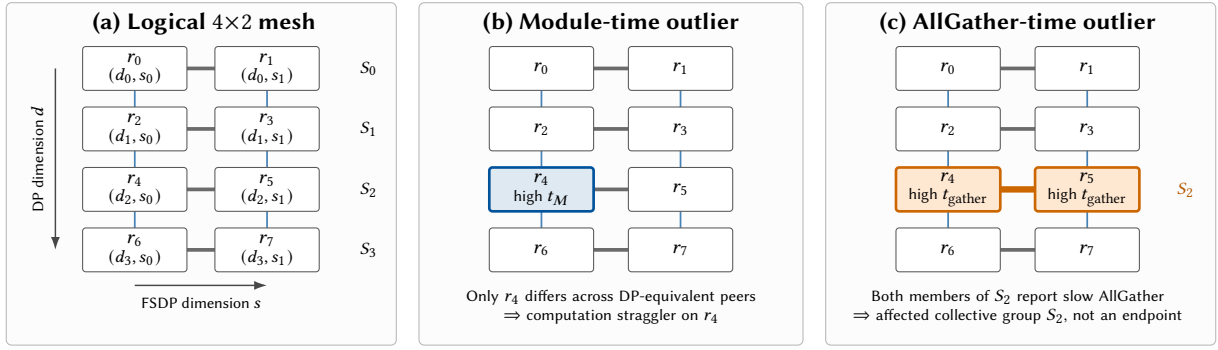

\noindent \textbf{Identical inputs.}
A module call may pass tensor inputs by position, such as a hidden state, or by name, such as an attention mask.
\system broadcasts these tensors from one source rank (Line 6), then applies exact C3 to the received inputs and RNG state (Line 7) to verify that every peer starts replay from identical values.
Non-tensor arguments remain local and must already agree.
For backward replay, the source broadcasts incoming output gradients, and each peer uses \texttt{autograd.grad} to return diagnostic input and parameter gradients without changing the training states.
For optimizer replay, the source broadcasts the copied pre-update parameter slice, effective gradient, optimizer state, and update configuration.
The source values need not be correct because spatial comparison requires only identical starting values across peers.

\noindent \textbf{Deterministic execution.}
Comparable replay of stochastic operators requires a common random stream and deterministic kernels.
\system broadcasts the source peer's pre-forward CPU, accelerator, and framework-specific model-parallel RNG states in $\rho_r$ with the input (Line 6), so healthy peers reproduce the same stochastic masks and outputs~\cite{pytorch-reproducibility-docs26,megatron-rng-docs26}.
After replay, each peer restores its saved state, so diagnosis neither advances the training RNG streams nor requires deterministic training.

\noindent \textbf{Input variation.}
Value variation broadens SDC coverage without changing the fixed tensor shapes.
One captured invocation tests only one value range, yet SDC-defective GPUs can fail only for particular training inputs~\cite{sdchunter-osdi26}.
Periodic capture changes activations, gradients, and parameters as training progresses.
Optionally, transforms such as $\{0.1x,x,10x\}$ widen the numerical ranges of floating-point inputs before Line 6 broadcasts them.
Each variant uses the same RNG state across peers, and C3 keeps output and gradient signatures separate rather than concatenating their tensors.

A complete layer recipe applies exact C3 to its output and backward gradients, and statistical C3 separately to every timed communication or computation interval.
After the real optimizer update, the optimizer recipe repeats the update on a copied slice and compares the results through exact C3, avoiding a copy of the complete model or optimizer.

\noindent \textbf{System overhead.}
Sampling one hidden layer at a configurable cadence makes the per-rank layer-equivalent replay estimate independent of the job's rank count.
For a model with $N$ repeated hidden layers, replaying one input variant through one hidden layer's forward and backward passes costs roughly $1/N$ of one training iteration time.
If \system replays $V$ input variants once every $I$ iterations, the amortized layer-equivalent overhead is therefore approximately $V/(IN)$.
For $V=3$, $I=20$, and $N=50$, this estimate is $3/(20\times50)=0.3\%$.

\subsubsection{Separating computation and communication}
\label{sec:sep-comp-comm}
\begin{figure}[!t]
\centering
\begin{tikzpicture}[
  font=\sffamily\scriptsize,
  panel/.style={draw=black!28, rounded corners=1.5pt, fill=black!1},
  gpu/.style={draw=black!65, rounded corners=1pt, fill=white, minimum width=0.72cm, minimum height=0.37cm, align=center, inner sep=1pt},
  fsdprank/.style={draw=orange!80!black, rounded corners=1pt, fill=orange!17, line width=0.9pt, minimum width=0.72cm, minimum height=0.37cm, align=center, inner sep=1pt},
  tprank/.style={draw=purple!80!black, rounded corners=1pt, fill=purple!12, line width=0.9pt, minimum width=0.72cm, minimum height=0.37cm, align=center, inner sep=1pt},
  candidate/.style={draw=red!75!black, rounded corners=1pt, fill=red!16, line width=1.3pt, minimum width=0.72cm, minimum height=0.37cm, align=center, inner sep=1pt},
  fsdp/.style={draw=black!58, line width=1.0pt},
  cthree/.style={draw=revisionblue!55, line width=0.55pt},
  slowfsdp/.style={draw=orange!80!black, line width=2.1pt},
  slowtp/.style={draw=purple!80!black, line width=2.1pt},
  axis/.style={-{Latex[length=1.4mm,width=1.0mm]}, draw=black!72, line width=0.5pt},
  result/.style={draw=red!75!black, rounded corners=1.5pt, fill=red!7, line width=0.8pt, text width=7.15cm, align=center, inner sep=2.5pt, font=\sffamily\scriptsize}
]

\draw[panel] (-3.90,0.70) rectangle (-0.15,-2.92);
\draw[panel] (0.15,0.70) rectangle (3.90,-2.92);
\node[font=\sffamily\scriptsize\bfseries] at (-2.02,0.47) {TP slice 0};
\node[font=\sffamily\scriptsize\bfseries] at (2.02,0.47) {TP slice 1};
\draw[axis] (-2.52,0.97) -- (2.52,0.97);
\node[font=\sffamily\tiny, fill=white, inner sep=1pt] at (0,0.97) {TP dimension $t$};

\draw[cthree] (-2.85,0.02) -- (-2.85,-2.14);
\draw[cthree] (-1.20,0.02) -- (-1.20,-2.14);
\draw[cthree] (1.20,0.02) -- (1.20,-2.14);
\draw[cthree] (2.85,0.02) -- (2.85,-2.14);

\foreach \yy in {0.02,-0.70,-2.14} {
  \draw[fsdp] (-2.85,\yy) -- (-1.20,\yy);
}
\draw[slowfsdp] (-2.85,-1.42) -- (-1.20,-1.42);
\foreach \yy in {0.02,-0.70,-1.42,-2.14} {
  \draw[fsdp] (1.20,\yy) -- (2.85,\yy);
}

\draw[slowtp] (-2.85,-1.23) -- (-2.85,-1.08) -- (1.20,-1.08) -- (1.20,-1.23);

\node[gpu] at (-2.85,0.02) {$r_0$};
\node[gpu] at (-1.20,0.02) {$r_1$};
\node[gpu] at (-2.85,-0.70) {$r_2$};
\node[gpu] at (-1.20,-0.70) {$r_3$};
\node[candidate] at (-2.85,-1.42) {$r_4$};
\node[fsdprank] at (-1.20,-1.42) {$r_5$};
\node[gpu] at (-2.85,-2.14) {$r_6$};
\node[gpu] at (-1.20,-2.14) {$r_7$};

\node[gpu] at (1.20,0.02) {$r_8$};
\node[gpu] at (2.85,0.02) {$r_9$};
\node[gpu] at (1.20,-0.70) {$r_{10}$};
\node[gpu] at (2.85,-0.70) {$r_{11}$};
\node[tprank] at (1.20,-1.42) {$r_{12}$};
\node[gpu] at (2.85,-1.42) {$r_{13}$};
\node[gpu] at (1.20,-2.14) {$r_{14}$};
\node[gpu] at (2.85,-2.14) {$r_{15}$};

\node[font=\sffamily\tiny, text=revisionblue, rotate=90] at (-3.38,-1.06) {C3-equivalent replicas};
\draw[axis] (-3.57,0.02) -- (-3.57,-2.14);
\node[font=\sffamily\tiny, rotate=90, anchor=south] at (-3.62,-1.06) {DP dimension $d$};
\draw[axis] (-2.85,-2.52) -- (-1.20,-2.52);
\node[font=\sffamily\tiny, anchor=north] at (-2.02,-2.56) {FSDP dimension $s$};

\node[result] at (0,-3.34) {Slow FSDP $\{r_4,r_5\}$ $\cap$ slow TP $\{r_4,r_{12}\}$ $= r_4$ $\Rightarrow$ replace the machine hosting $\{r_4,r_{12}\}$};

\end{tikzpicture}
\caption{Cross-PG validation localizes the faulty machine: the slow FSDP and TP groups intersect at $r_4$, so \system selects the machine hosting $\{r_4,r_{12}\}$ for replacement.}
\Description{A single-column figure shows sixteen GPUs in two side-by-side tensor-parallel slices, each containing a four-by-two data-parallel and FSDP mesh.
Thin blue vertical lines connect C3-equivalent replicas but do not represent slow communication groups.
An orange horizontal line highlights the slow FSDP group containing ranks r4 and r5, and a purple line across the two slices highlights the slow TP group containing ranks r4 and r12.
Rank r4 is outlined in red because it is the only rank common to the two slow communication groups.
A box below the meshes states that this intersection localizes the machine hosting the tensor-parallel group containing ranks r4 and r12 as the replacement unit.}
\label{fig:cross-pg-straggler-localization}
\end{figure}
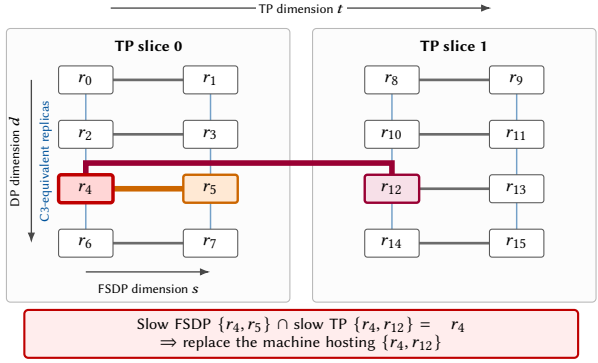

A single replay duration cannot distinguish slow module computation from time spent in collective communications, yet the two cases require different localization evidence.
\system therefore times computation and communication separately.
Lines 8--9 in Algorithm~\ref{alg:dense-replay} separately record $t_{\mathrm{gather}}$ for parameter AllGather and $t_M$ for module computation.
Lines 10--13 then compare the outputs of both the collective communication and the module computation with exact C3 and statistical C3.
An outlier in $t_M$ localizes the slowdown to module $M$ and an outlier in $t_{\mathrm{gather}}$ instead identifies FSDP parameter communication as the affected surface.

Figure~\ref{fig:compute-communication-separation} shows why the two timings yield different spatial evidence.
Figure~\ref{fig:compute-communication-separation} organizes eight ranks as a $4\!\times\!2$ DP--FSDP mesh, where each row is an FSDP shard PG and each column contains C3-equivalent replicas.
In Figure~\ref{fig:compute-communication-separation}(b), only $r_4$ has an elevated $t_M$, so C3 directly identifies $r_4$ as the computation straggler.
In Figure~\ref{fig:compute-communication-separation}(c), both members of $S_2=\{r_4,r_5\}$ observe an elevated $t_{\mathrm{gather}}$ because the AllGather synchronizes them.
The communication timing therefore identifies the affected PG $S_2$, but not which member caused the slowdown.

\system uses cross-PG validation to convert affected PGs into an actionable machine-level diagnosis.
Practical LLM pre-training combines multiple parallelism strategies and confines each TP group within one machine because TP communicates heavily~\cite{megatron-arxiv19,megascale-nsdi24}.
Each rank consequently participates in multiple PGs, allowing abnormal PGs from different parallelism dimensions to identify their common endpoint and its host machine.

Figure~\ref{fig:cross-pg-straggler-localization} extends Figure~\ref{fig:compute-communication-separation}'s $4\!\times\!2$ DP--FSDP mesh with TP degree two, yielding 16 ranks arranged in two TP slices.
Ranks with the same DP and FSDP coordinates across the slices form an intra-machine TP group, while the blue DP-aligned columns remain C3 comparison groups.
In the example, the slow FSDP group $\{r_4,r_5\}$ and slow TP group $\{r_4,r_{12}\}$ intersect only at $r_4$.
Because $\{r_4,r_{12}\}$ occupies one machine, \system localizes that machine for replacement rather than reporting $r_4$ as the recovery unit.

The joint TP--FSDP latency pattern separates four high-level diagnoses.
Slow TP and intersecting FSDP PGs identify a machine-local problem.
A slow TP PG with normal FSDP latency identifies an intra-machine communication problem on the TP group's machine.
Normal TP latency with slow FSDP PGs indicates an inter-machine communication problem; \system reports the affected PGs and leaves link- or device-level localization to existing datacenter network-diagnosis systems~\cite{netbouncer-nsdi19,passive-network-nsdi17}.
Only when many independent PGs spanning multiple machines slow simultaneously and no single machine explains them does \system report a job-level software issue or cluster-wide contention.
The same principle applies to other hybrid-parallel configurations whenever independently observed PGs provide overlapping membership evidence for the same machine.

\subsection{MoE-layer replay}
\label{sec:moe-shape-coverage}

Compared with a dense Transformer block, an MoE layer adds three stages: a router assigns tokens to experts, an AllToAll exchanges routed tokens across expert-parallel ranks, and expert computation processes the received tokens~\cite{gshard-arxiv20,olmoe-arxiv24,megascale-moe-eurosys26}.
For a fixed batch size and sequence configuration, the router still receives a fixed-shape input, so the dense replay rules apply to it.

\noindent \textbf{AllToAll communication.}
\system separates AllToAll diagnosis by failure manifestation.
Replay targets SDC and software- or hardware-induced stragglers, while \system's OOB observers in Section~\ref{sec:oob-hangs} handle collective hangs.
Routing can legitimately produce uneven token splits and slow a healthy collective, so workload-induced imbalance is not itself failure evidence~\cite{stragglers-osdi25,sabre-ics26}.
To separate targeted failures from this workload variation, corresponding EP positions in independent replicas execute the same communication plan.
As in Algorithm~\ref{alg:dense-replay}, C3 compares their numerical outputs and durations to expose numerical and timing outliers.
GPUs hosting experts also participate in process groups from other parallelism dimensions, such as TP or DP.
Cross-PG validation (\S~\ref{sec:sep-comp-comm}) can therefore intersect slow groups to narrow the fault to a shared machine or communication endpoint.
\system also generates representative traffic matrices for AllToAll replay to broaden coverage across dynamic input shapes.

\noindent \textbf{Shape-dependent failures in Expert computation.}
Routing determines the number of tokens processed by each expert, which is one dimension of the expert's input shape.
Changing this token count changes the amount of computation and can change how a grouped kernel schedules it~\cite{megablocks-arxiv22,nvidia-cutlass-grouped26}.
Because production SDCs can depend on the input data and the affected GPU functional unit, while grouped-kernel latency depends on scheduling, a numerical error or slowdown may appear at one expert-input shape but not another~\cite{sdchunter-osdi26,nvidia-cutlass-grouped26}.
Replaying only the shape observed in one layer of one iteration can therefore miss a recurring failure that appears under another routed token distribution.
Exhaustively replaying every admitted shape online is also impractical because the search space can contain thousands of shapes, increasing diagnosis latency and runtime overhead.

\noindent \textbf{Opportunity.}
\system considers shape compression only after verifying that candidate shapes select the same kernel and scheduling configuration.
A GPU GEMM divides its output matrix into tiles, and a grouped-kernel scheduler assigns those tiles to thread blocks~\cite{nvidia-cutlass-grouped26}.
An \emph{execution path} is a distinct sequence of kernel operations used to process one scheduled unit of work, such as an interior tile, a boundary tile, or a tail tile.
A single input shape can exercise several execution paths and may execute each path many times.
Suppose shape $\alpha$ executes every qualified path at least as many times as another shape $\beta$ with the same kernel and scheduling configuration.
Any persistent fault that appears after a particular number of executions under $\beta$ can also appear under $\alpha$ because $\alpha$ executes the same execution path at least that many times.
\system can therefore retain $\alpha$ and omit $\beta$ from the catalog.

\noindent \textbf{Shape coverage.}
\system compresses the search space of shape-dependent replay by omitting shapes whose execution paths are already covered by other shapes.
The \emph{execution fingerprint} $F(\alpha)$ summarizes shape $\alpha$'s kernel and scheduling configuration.
The path-count vector $\mathbf{c}(\alpha)$ records how many times $\alpha$ exercises each execution path in that configuration.
When two fingerprints match, corresponding components of their path-count vectors refer to the same execution paths and can be compared directly.
Shape $\alpha$ covers shape $\beta$ when their fingerprints match and $\alpha$'s path-count vector is no smaller in any component:
\[
  \alpha \succeq \beta
  \quad\Longleftrightarrow\quad
  F(\alpha)=F(\beta)
  \;\land\;
  \mathbf{c}(\alpha)\ge\mathbf{c}(\beta)
  \quad\text{componentwise}.
\]
For example, if $\mathbf{c}(\alpha)=(10,1)$, and $\mathbf{c}(\beta)=(6,1)$, then $\alpha$ covers $\beta$.
For each candidate $\alpha$, the set of shapes it can cover is
\[
  \mathcal{C}(\alpha)=\{\beta\mid\alpha\succeq\beta\}.
\]
A \emph{representative} is a retained shape selected to cover part of the admitted set.
\system retains the fewest representatives whose coverage sets together contain every admitted shape.
If \system cannot determine an execution fingerprint or path-count vector reliably, it disables compression and retains every shape as a representative.

\noindent \textbf{Offline discovery.}
\system builds the representative catalog before training so exhaustive profiling does not consume training resources or lengthen online diagnosis.
Offline discovery is valid only for a fixed training environment, so each catalog is bound to the GPU type, software stack, and model configurations that determine kernel selection.
Any environment drift invalidates the catalog and requires rediscovery.
To build the catalog, \system enumerates the admitted expert-input shapes for each represented layer, expert, and expert-parallel position and runs the production expert implementation.
\system records the execution fingerprint and path-count vector of each shape; it then derives the minimum set of representatives that cover the entire search space.

\noindent \textbf{Online replay.}
Online replay rotates through the offline representatives instead of testing every admitted shape at once.
At each scheduled check, \system selects the next representative and constructs identical inputs of that shape for equivalent peers.
The peers replay the expert computation, and C3 compares their outputs and gradients to localize any disagreement.
After a successful check, \system advances to the next representative until the catalog is covered.
We call one full rotation through the $K$ representatives a \emph{recipe cycle}.
With replay interval $I$, a recipe cycle takes at most $KI$ training steps and supplies the evidence required for checkpoint certification in Section~\ref{sec:checkpoint-certification}.

\section{Out-of-Band Hang Localization}
\label{sec:oob-hangs}

Hang localization requires evidence that remains available after the trainer or its collective communications stop making progress.
\system therefore separates progress observation from the training process and from its communicators.
The mechanism turns a job-wide timeout into either a rank-local progress divergence, a visible collective-protocol mismatch, or a group-scoped runtime stall.

\noindent \textbf{Why communication hangs.}
A collective timeout is a catch-all symptom rather than evidence that the communication library itself failed.
PyTorch's fleet analysis groups the common causes into four categories: (1) CPU-side stalls or control-flow divergence, (2) a preceding GPU-kernel hang, (3) incompatible collective arguments, and (4) network or hardware faults~\cite{flight-recorder26}.
At the watchdog, these causes appear in two ways: either the collective kernel itself does not finish before the timeout, or ranks have desynchronized and are no longer executing compatible collective operations.
The study reports that almost all observed timeouts arise from desynchronization, so the rank that reports the timeout and the collective visible at that instant may both be victims rather than causes.

\noindent \textbf{Comparable progress.} \system primarily targets hangs caused by desynchronization.
Framework hooks publish a monotonic training coordinate consisting of the optimizer step and an operation identifier at repeated-layer, DataLoader, and supported collective boundaries.
Immediately before a supported collective, \system also publishes a fingerprint of the process group, collective type, and arguments that healthy callers must share, including fixed shape, dtype, reduction, and root metadata.
Legitimately rank-dependent values, such as variable expert-parallel AllToAll splits, require a compatibility rule and are not compared by simple equality.
The progress schema records intent before a rank enters the operation that may block it.
A lagging rank therefore leaves a different coordinate, while ranks already waiting in the collective retain the later coordinate and pending fingerprint.
This ordering avoids relying on the rank that first reports the training-communicator timeout, which may be a healthy victim.

\noindent \textbf{An independent observer.}
Each rank launches a CPU-only observer process that reads the latest progress record from shared memory.
Observers communicate through a Gloo group created separately from the training communicator, so a blocked trainer and NCCL communicator do not directly block progress consensus.
The training path only updates local shared memory and signals a local event.
The observer enters C3 only after visible progress remains silent beyond a configured threshold.
This event-triggered design avoids periodic diagnostic collectives during healthy execution while ensuring that the observers of synchronously blocked peers eventually enter the same comparison.

\noindent \textbf{Guiding recovery.}
The OOB classification guides the choice between restarting the job on the same resources and inspecting or replacing hardware before restart.
When a minority of ranks publishes a different collective fingerprint, \system attributes the hang to rank-local software or control-flow divergence.
Recovery can then correct the protocol error and restart the job on the same hardware.
When all ranks report matching progress and fingerprints, \system instead classifies a group-scoped runtime or transport stall.
Recovery then runs device and fabric diagnostics and replaces a resource only if they confirm a hardware fault.

\section{Certifying Recovery Checkpoints}
\label{sec:checkpoint-certification}

A latent SDC can corrupt model or optimizer state and contaminate recent checkpoints before replay exposes it, so restarting from the newest checkpoint can reintroduce corrupted state.
A checksum verifies that stored bytes match the captured bytes, but cannot establish that the captured training state was numerically correct.
\system couples replay evidence to checkpoint eligibility and resumes training jobs only from certified checkpoints that preserve numerical trust.

To reduce recovery latency, \system pairs with \gemini~\cite{gemini-sosp23}, which asynchronously copies post-optimizer training state from GPU to CPU memory for frequent capture and fast retrieval.
Because dense models have static shapes during training, every in-memory checkpoint captured after an accepted \system check is certified by that check.

However, MoE routing produces dynamic replay shapes, so one accepted check does not cover the entire recipe catalog.
\system therefore persists the latest rank-consistent \gemini checkpoint as a durable candidate only after a complete cycle supplies accepted evidence for all $K$ recipes.
The checkpoint protocol for MoE models tracks three checkpoint states: \emph{latest}, \gemini's newest complete in-memory checkpoint; \emph{candidate}, the persisted checkpoint awaiting complete validation; and \emph{verified}, the checkpoint whose numerical trust is supported by \system's replay evidence.

\begin{algorithm}[!t]
\caption{Checkpoint saving and retrieval}
\label{alg:checkpoint-save}
\KwIn{Accepted-recipe count $q=0$; recipe-catalog size $K$; latest in-memory checkpoint $CKPT_{latest}$; current candidate $CKPT_{candidate}$; newest verified checkpoint $CKPT_{verified}$}
\Fn{\textsc{CheckpointSave}$()$}{
  {$f \gets$ job-wide \system outcome\;}
  \If{$f \in \{\textsc{SDC},\textsc{Hang}\}$}{
    {\Return \textsc{CheckpointRetrieval}(f)\;}
  }
  {$CKPT_{latest} \gets$ newly captured in-memory checkpoint\;}
  {$q \gets q+1$\;}
  \If{$q=K$}{
    {$q \gets 0$\;}
    {$CKPT_{verified} \gets CKPT_{candidate}$\;}
    {persist $CKPT_{latest}$\;}
    {$CKPT_{candidate} \gets CKPT_{latest}$\;}
  }
}

\Fn{\textsc{CheckpointRetrieval}$(f)$}{
  \If{$f=\textsc{SDC} \lor \neg\textsc{MachinesAccessible}()$}{
    {\Return $CKPT_{verified}$\;}
  }
  \If{$f=\textsc{Straggler}$}{
    {\Return $CKPT_{latest}$\;}
  }
  {$R \gets \textsc{ReplayFullCatalog}()$\;}
  \If{$\textsc{HasSDC}(R)$}{
    {\Return $CKPT_{verified}$\;}
  }
  {\Return $CKPT_{latest}$\;}
}
\end{algorithm}

\noindent \textbf{Checkpoint saving during training.}
We define a \emph{replay boundary} as the point immediately after the optimizer update in an iteration selected by \system's configured replay cadence.
At every replay boundary, \system invokes \textsc{CheckpointSave()} in Algorithm~\ref{alg:checkpoint-save}; its out-of-band observer enters the same procedure when it detects a communication hang.
An SDC reported by any comparison group or a communication hang immediately invokes \textsc{CheckpointRetrieval()}.
A healthy result or completed straggler instead permits \gemini to capture the current state and increments $q$, the number of accepted recipes in the current cycle.
\system reports the straggler separately because slowness alone is not numerical divergence.
When $q$ reaches $K$, \system resets the counter, promotes the prior candidate to \emph{verified}, and persists the latest \gemini checkpoint as the new candidate.
One cycle is insufficient because a fault can corrupt training state after its triggering recipe has already run.
The checkpoint captured at that cycle boundary can therefore contain corruption that the first cycle did not test.
The next cycle replays all $K$ recipes after capture, allowing \system to expose such corruption before promotion.
Thus, \system promotes a candidate only after two consecutive accepted cycles, comprising $2K$ checks.

\noindent \textbf{Checkpoint retrieval for failure recovery.}
The \textsc{CheckpointRetrieval()} procedure in Algorithm~\ref{alg:checkpoint-save} uses the failure class, machine accessibility, and, when necessary, the full-catalog replay result to choose between the latest and verified checkpoints.
A confirmed SDC selects the verified checkpoint.
If a machine is disconnected, \system cannot inspect its latest shard or establish that the latest job-wide checkpoint is numerically clean, so it also selects the verified checkpoint.
A straggler selects the latest \gemini checkpoint because slowness alone does not indicate numerical corruption.
For any other failure, \system runs the complete compressed recipe catalog and selects the latest checkpoint only if the sweep completes without SDC; otherwise, it selects the verified checkpoint.
The emergency sweep replays only the selected layers and captured inputs; it neither restores nor re-executes the full model checkpoint.
\gemini retrieves the selected checkpoint for newly added machines that replace faulty ones~\cite{gemini-sosp23}.

\section{Implementation}
\label{sec:implementation}

\system is implemented as a Python library over PyTorch~\cite{pytorch-neurips19}.
Applications invoke \texttt{enable\_resiliency} with the model, optimizer, and optional replay, checkpoint, and parallelism configuration.
Framework adapters discover repeated layers, optimizer wrappers, and replica groups, then install public module and optimizer hooks without modifying the training loop or framework source.
Adapters support TorchTitan, Megatron-Core~\cite{megatron-arxiv19}, and DeepSpeed with ZeRO~\cite{zero-sc20}.

\noindent \textbf{Structured replay.}\hspace{0.35em}
Forward and backward hooks retain the sampled layer's pytree arguments, tensor-leaf gradient outputs, and input \texttt{requires\_grad} mask.
Dynamic-shape rounds clone captured tensors; other rounds retain detached views only through the optimizer boundary.
\system broadcasts tensor leaves and reconstructs the pytree, while adapters must make non-tensor leaves equivalent.
Replay temporarily installs the source peer's CPU, CUDA, and Megatron model-parallel RNG states, scopes PyTorch deterministic-algorithm mode to the replay, and restores local state afterward.
\system computes replay gradients with \texttt{autograd.grad}.
It never passes live parameter gradients to \texttt{backward}.
TorchTitan materializes only the sampled FSDP2 layer; HSDP first compares corresponding local shards, then replays training-shaped ReduceScatter and replica AllReduce with scratch tensors.
Pure FSDP omits shard-correctness checks because no corresponding owner copy exists.

\noindent \textbf{Optimizer boundary.}\hspace{0.35em}
For replicated PyTorch optimizers, \system compares sampled weights after the real \texttt{optimizer.step}; DeepSpeed and Megatron adapters hook their underlying PyTorch \texttt{Optimizer}.
On a scheduled check, the pre-hook clones a rotating 64K-element slice of the parameter, effective gradient, group configuration, optimizer slots, and optimizer-owned tensors.
After the real step runs unchanged, \system applies the same base optimizer to the clone and compares the two updated slices.
This surface requires corresponding replicated ownership and remains disabled for uniquely owned sharded updates.

\noindent \textbf{Progress path and communication.}\hspace{0.35em}
Each rank launches a separate CPU observer.
Training hooks publish step, operation ID, pending-collective fingerprint, and sampled DataLoader timing to a lock-free shared-memory record and signal visible progress.
The observers communicate through an independently rendezvoused Gloo group, so a blocked training communicator does not directly block hang localization.
Replay uses a DP-aligned Gloo group for scalar and object metadata and a separate NCCL group for tensor broadcast and signature exchange.
\system folds contiguous tensor bytes to at most 512 64-bit values on the source device, finalizes one 64-bit signature on the CPU, and AllGathers only signatures; multiple numerical surfaces share one exchange but retain separate outlier bitmaps.
Reports preserve peer ranks, per-surface SDC bitmaps, replay timing, layer, mode, shape, progress, collective fingerprints, stall duration, and attribution scope so recovery policy can distinguish rank outliers from group-scoped stalls.

\noindent \textbf{Recovery integration.}\hspace{0.35em}
\system's checkpoint coordinator converts replay results into \gemini~\cite{gemini-sosp23} checkpoint operations.
Before mutating checkpoint state, it combines missing-result, SDC, and recipe-cycle flags across the job; incomplete evidence skips capture, and disagreement about a cycle boundary raises an error.
At a cycle boundary, \gemini completes any peer replication, verifies through a collective minimum that every checkpoint rank holds the same generation, and writes each local shard together with its received peer replica when explicit replication is enabled.
Loading performs the same rank-consistency check before selecting an in-memory or persisted generation.
Per-shard status metadata, temporary-file writes, \texttt{fsync}, and atomic replacement preserve the selected recovery role across restart.

\section{Evaluation}
\label{sec:evaluation}

We evaluate \system by asking three questions: (1) Can \system localize different types of latent failures; (2) does its evidence prevent recovery from selecting a contaminated checkpoint; and (3) how much can qualified MoE execution regimes compress replay?
The experiments answer these questions on a focused two-host testbed and dedicated qualification harnesses.

\subsection{Methodology and small-testbed setup}

\noindent \textbf{Hardware and software.}\hspace{0.35em}
Our main testbed comprises two hosts, each with eight NVIDIA A100-SXM4-40GB GPUs, for 16 training ranks.
NVLink connects GPUs within a host, while inter-host NCCL and Gloo traffic uses TCP since the testbed has no RDMA devices.
The framework campaign uses PyTorch 2.13.0 with CUDA 13.0, NCCL 2.29.7, TorchTitan 0.2.2, Megatron-Core 0.18.2, and DeepSpeed 0.19.4.

\noindent \textbf{Workloads.}\hspace{0.35em}
The experiments train a deterministic three-block Transformer with AdamW under DDP, FSDP2, and a $4\!\times\!4$ HSDP mesh.
The broader integration matrix exercises framework-native pipeline, tensor, context, sequence, data, and expert parallelism, including Megatron/Transformer Engine and DeepSpeed MoE layers with routed forward and backward execution.

\subsection{Fault-injection and recovery effectiveness}

\begin{table*}[!t]
\centering
\small
\begin{tabular}{@{}>{\raggedright\arraybackslash}p{0.15\textwidth}>{\raggedright\arraybackslash}p{0.38\textwidth}>{\raggedright\arraybackslash}p{0.39\textwidth}@{}}
\toprule
Fault & Injected trigger & Observed coverage \\
\midrule
Dense SDC & Rank-9 DDP parameter after backward, FSDP2 replay output, or HSDP gradient shard before the optimizer step & Exact rank in 3/3 cells; contaminated boundary excluded and prior state restored bitwise \\
Dense numerical SDC & Forward-output or parameter perturbation, including 64 near-invisible cases & 344/344 localized; 30/30 fault-free runs stayed clean \\
Compute straggler & Rank-9 sampled-layer forward delayed by 250\,ms & Exact rank in 3/3 cells after two confirmations; fault-free replay below 1.6\,ms \\
Communication straggler & Rank-15 tensor-parallel collective delayed & Affected rank-14/15 group localized; no compute-straggler or SDC report; next step clean \\
MoE SDC & Persistent rank-15 expert-weight corruption at 128- and 512-row routed shapes & Exact rank at both shapes after one clean recipe cycle \\
Hang and input stall & Frozen progress, divergent operation metadata, or delayed input fetch & 120/120 primary and 30/30 focused launches passed \\
MoE kernel-role SDC & Every selected semantic work-item occurrence perturbed & 5,960/5,960 changed output or gradient; persistent faults localized ranks 7 and 15 on 8 and 16 GPUs \\
\bottomrule
\end{tabular}
\caption{Fault-injection coverage with 16-GPU training jobs.}
\label{tab:fault-injection-coverage}
\end{table*}

\noindent \textbf{Protocol and metrics.}\hspace{0.35em}
SDC tests run three steps: one with injection disabled, one with a fault at the scheduled replay boundary, and one after disabling the injector again.
The injection-free first step must produce no false positive; after the one-step fault is removed, the third step must again produce no SDC report, showing that \system no longer flags the rank.
Checkpoint-recovery tests instead run two accepted checks before injecting persistent SDC, while compute-straggler tests require the same rank in two consecutive timing rounds.
We record false positives, the reported rank or group, fault class, checkpoint decision, and whether a fresh runtime restores the selected state bitwise.

\noindent \textbf{Fault model.}\hspace{0.35em}
Table~\ref{tab:fault-injection-coverage} groups injections by manifestation and experiment purpose.
The dense SDC row couples an architecture-specific numerical corruption to checkpoint exclusion and restart, while the dense numerical SDC row applies a broader replay-only perturbation matrix.
A compute straggler delays sampled-layer execution, whereas a communication straggler delays a collective.
MoE SDC corrupts expert state during routed training, while MoE kernel-role SDC perturbs selected semantic work inside the instrumented kernel.
Hang and input-stall cases stop visible progress through a frozen process, incompatible collective metadata, or delayed input fetching.

Table~\ref{tab:fault-injection-coverage} reports measured coverage for each modeled injection.
The current two-host training-job cells use one deterministic seed and inject rank 9 or 15, so their success fractions cover only the stated injections, not physical failures in general.
These software injections validate the mechanism and its stated localization scope, not recall over arbitrary physical failures.

The current checkpoint protocol recovers the correct state in the tested fault cells.
Across DDP, FSDP2, and HSDP, all nine 16-GPU cells for checkpoint recovery, SDC exclusion, and compute-straggler localization passed.
Each SDC-contaminated boundary was excluded globally, and a fresh runtime restored the preceding model, AdamW, caller-owned, CPU-RNG, and CUDA-RNG state bitwise.
An eight-GPU, two-recipe test further made step 2 a candidate, promoted it after the clean check at step 4, injected SDC at step 5, and recovered step 2.
These results establish recovery-state correctness for the injected scenarios.

\subsection{MoE replay coverage and compression}

We profile an instrumented BF16 Triton grouped-GEMM on A100 GPUs, executing every admitted expert row count three times and requiring a stable kernel fingerprint.
The compressor retains representatives that jointly cover every recorded semantic work-item count within each execution and pressure regime.
Single-expert experiments vary the projection dimensions; grouped experiments fix each expert projection at $128\!\times\!128$, assign the same row count to every physically executed expert, and vary the number of experts per kernel.
In Table~\ref{tab:moe-catalog-size}, ``admitted'' counts exhaustively profiled row counts and ``selected'' counts the compressed replay plans.

\begin{table}[!t]
\centering
\small
\begin{tabular}{@{}lrrr@{}}
\toprule
Configuration & Admitted & Selected & Reduction \\
\midrule
\multicolumn{4}{@{}l}{\itshape Single-expert projections} \\
\midrule
$128\!\times\!128$ & 3,457 & 16 & 99.54\% \\
$6144\!\times\!10752$ & 2,048 & 10 & 99.51\% \\
$5120\!\times\!1536$ & 1,024 & 18 & 98.24\% \\
$2048\!\times\!1024$ & 2,048 & 19 & 99.07\% \\
\midrule
\multicolumn{4}{@{}l}{\itshape Grouped kernel, uniform $128\!\times\!128$ expert GEMMs} \\
\midrule
$2$ experts per kernel & 2,048 & 18 & 99.12\% \\
$4$ experts per kernel & 2,048 & 27 & 98.68\% \\
$8$ experts per kernel & 2,048 & 42 & 97.95\% \\
$16$ experts per kernel & 2,048 & 48 & 97.66\% \\
\bottomrule
\end{tabular}
\caption{Coverage-based compression in MoE replay.}
\label{tab:moe-catalog-size}
\end{table}

Table~\ref{tab:moe-catalog-size} shows that the exhaustive $128\!\times\!128$ campaign compressed 3,457 admitted shapes into 16 representatives, while the other single-expert projections retained 10--19.
For uniform grouped execution with two to 16 experts per kernel, the 2,048-shape catalogs retained 18--48 representatives.
The growth reflects new pressure regimes and non-dominated work-item counts as each kernel executes more experts.
The grouped campaigns passed 4,757 selected execution-path injections, and host-swapped repetitions reproduced their catalogs.
The 16 single-expert representatives covered all 1,134,224 recorded execution-path occurrences across the admitted shapes, and all 5,960 injected selected-role occurrences changed the forward output, input gradient, or weight gradient.
These grouped results cover uniform per-expert row counts; arbitrary heterogeneous routing vectors require separately qualified templates.

\section{Discussion and Limitations}

\noindent \textbf{Evaluation scale.}\hspace{0.35em}
The current evaluation exercises software fault injection and dense and MoE mechanisms on at most 16 A100 GPUs across two hosts.
A production-scale study should additionally quantify detection accuracy, false positives, and time to evidence across multiple racks, hybrid-parallel layouts, and RDMA or multi-rail fabrics, including cross-process-group endpoint localization.
It does not measure end-to-end throughput and resource overhead across replay cadences, or report recovery time and rollback distance.
Paired in-situ and offline diagnosis and ablations of majority consensus, temporal confirmation, the independent OOB observer, controlled replay, and catalog compression remain future work.
The present results therefore establish mechanism behavior for the stated injections and configurations, while comparative effectiveness, production-scale cost, and end-to-end recovery savings remain to be measured.

\noindent \textbf{Replay coverage.}\hspace{0.35em}
Replay covers the layers, values, shapes, and communication paths exercised by its rotating checks.
\system targets permanent faults and intermittent faults that recur under a similar operating regime; a one-shot fault that does not recur during replay falls outside this coverage contract.
MoE catalog compression is qualified for a particular backend, hardware and software environment, and admitted shape domain.
The current measurements cover uniform expert row counts, while arbitrary heterogeneous routing vectors require separately qualified templates.
For dynamic expert-parallel AllToAll, \system claims compressed coverage only if profiling shows that the admitted traffic matrices map to a bounded set of communication execution paths, physical resources, and load regimes; the present evaluation does not establish that condition.

\noindent \textbf{Diagnosis scope.}\hspace{0.35em}
\system localizes an actionable rank, GPU, node, peer group, or conditionally an HCA/NIC endpoint.
Identifying a lower-level cause, such as a kernel instruction, cable, port, or switch, requires external component telemetry or fabric diagnostics.
\system's temporal reference detects equal-progress hangs through a progress timeout and peer-group-wide slowdowns relative to clean history, although either symptom requires additional evidence for source attribution.
Moreover, public framework interfaces do not expose every collective launched inside FSDP, DTensor, fused kernels, or compiled graphs.
Version-specific launch-boundary adapters can extend this visibility; without one, \system preserves a group-scoped diagnosis, and the current implementation does not automatically time every parallel process group.

\section{Related Work}

\noindent \textbf{End-to-end training resilience.}
ByteRobust~\cite{byterobust-sosp25} combines lifecycle monitoring, fault isolation, replay, warm standbys, hot updates, and high-frequency checkpointing to sustain production LLM training.
TrainMover~\cite{trainmover-osdi26} instead prepares elastic or standby machines through delta-based communication-group setup and sandboxed warmup, reducing downtime when an interrupted job changes membership.
Check-N-Run~\cite{checknrun-nsdi22}, \gemini~\cite{gemini-sosp23}, and Bamboo~\cite{bamboo-nsdi23} provide complementary checkpointing and fault-tolerance mechanisms for restoring execution after known failures.
\system supplies culprit and checkpoint-eligibility evidence for latent failures, which complements these systems' resource replacement, reconfiguration, and state-loading mechanisms.

\noindent \textbf{Hangs and stragglers.}
Mycroft~\cite{mycroft-sosp25} traces internal collective-communication states and reconstructs control and data dependencies to localize failures and stalls that coarse framework traces hide.
The OSDI straggler study~\cite{stragglers-osdi25} uses counterfactual what-if analysis over production traces to quantify straggler impact and investigate temporal, spatial, and causal patterns.
Minder~\cite{minder-nsdi25} and Aegis~\cite{aegis-nsdi25} use job-level telemetry and diagnostic rules to detect faulty machines.
Holmes~\cite{holmes-nsdi25}, GREYHOUND~\cite{greyhound-atc25}, and EROICA~\cite{eroica-nsdi26} use cross-layer performance evidence to localize training slowdowns.
\system complements historical and collective-internal diagnosis with a controlled rank differential: its OOB observers compare progress and collective fingerprints after a trainer blocks, while in-situ replay compares equivalent computation and communication among live peers.

\noindent \textbf{Infrastructure and component diagnosis.}
NetBouncer~\cite{netbouncer-nsdi19} actively probes datacenter paths to localize failed devices and links, while passive network diagnosis combines path information with end-host statistics to localize partial network failures~\cite{passive-network-nsdi17}.
For accelerators, DCGM exposes GPU health telemetry and active checks~\cite{dcgm-docs26}, while SuperBench runs controlled validation workloads across AI infrastructure~\cite{superbench-tocs26}.
For CPUs and host memory, Linux RAS exposes machine-check and EDAC reports for processor, cache, and memory errors~\cite{linux-ras-docs26}.
Linux In-Field Scan additionally runs circuit-level tests on CPU cores for defects that parity and ECC do not detect~\cite{linux-ifs-docs26}.
These mechanisms identify a replaceable component when a counter, probe, or diagnostic workload exposes the fault.
\system addresses complementary latent cases whose training symptom lacks such an authoritative component signal; its rank- or peer-group verdict narrows the endpoint that infrastructure tools can probe before remediation.

\noindent \textbf{Numerical correctness and SDC diagnosis.}
TrainCheck~\cite{traincheck-osdi25} infers and checks training invariants online, while TrainVerify~\cite{trainverify-sosp25} formally verifies a distributed plan's equivalence to the model specification.
AEGIS~\cite{aegis-sdc-osdi26} separates lightweight corruption sensing from definitive verification; SDCHunter~\cite{sdchunter-osdi26} replays the triggering workload to isolate a defective GPU and resume from a verified checkpoint.
OpGuard~\cite{opguard-osdi26} localizes the first bitwise operator mismatch, Dr. DNA~\cite{drdna-asplos24} checks activation distributions inline, and SuperBench~\cite{superbench-tocs26} exercises infrastructure with controlled diagnostics.
\system derives a live reference by equalizing inputs and randomness across an equivalent peer group and applying strict-majority consensus to numerical signatures.
TrainCheck's semantic invariants and TrainVerify's plan equivalence validate complementary correctness properties; \system adds controlled peer comparison that preserves rank identity and binds clean replay coverage to recovery-state eligibility.
AEGIS~\cite{aegis-sdc-osdi26} and SDCHunter~\cite{sdchunter-osdi26} use SDC evidence to support rollback; \system ties completed, clean replay coverage to a candidate, promotes it after all required ranks agree, and points recovery to the newest checkpoint supported by that evidence.

Across these categories, \system connects progress, timing, and numerical evidence through one peer-consensus interface.
Its evidence complements recovery orchestration, collective-internal visibility, and production SDC diagnosis at different layers of the LLM training stack.

\section{Conclusion}

\system makes runtime diagnosis additive to distributed training.
Its OOB service remains available when a trainer blocks, while in-situ replay preserves the production model, kernels, allocations, and operating pressure.
Equivalent peers turn progress, timing, and numerical outputs into evidence for distinguishing protocol divergence, group-scoped stalls, and rank outliers; topology and majority evidence bound attribution.
Clean replay coverage also certifies the checkpoint eligible for recovery.
Together, graceful attachment, production-condition replay, conservative localization, and checkpoint certification form \system's central design.

\bibliographystyle{ACM-Reference-Format}
\bibliography{references}
\end{document}